\documentclass[a4paper,twocolumn,11pt,unpublished]{quantumarticle}
\pdfoutput=1
\usepackage[utf8]{inputenc}
\usepackage[english]{babel}
\usepackage[T1]{fontenc}
\usepackage{amsmath}

\usepackage[numbers]{natbib}
\usepackage{hyperref}
\usepackage{amssymb}
\usepackage{booktabs}
\usepackage{float}
\usepackage{subcaption}

\usepackage{braket}
\usepackage{multirow}

\usepackage{algorithm}
\usepackage{algpseudocode}

\usepackage{tikz}
\usepackage{lipsum}

\begin{document}

\title{SWAP-Network Routing and Spectral Qubit Ordering for MPS Imaginary-Time Optimization}
% SWAP-Network Routing and Spectral Qubit Ordering for MPS Imaginary-Time Optimization      

\author{Erik M. Åsgrim}
\affiliation{Division of Computational Science and Technology, KTH Royal Institute of Technology, Stockholm, Sweden}
\orcid{0009-0001-5695-209X}
\email{erima@kth.se}

\author{Stefano Markidis}
\affiliation{Division of Computational Science and Technology, KTH Royal Institute of Technology, Stockholm, Sweden}
\orcid{0000-0003-0639-0639}
\email{markidis@kth.se}

\maketitle

\begin{abstract}
We propose a quantum-inspired combinatorial solver that performs imaginary-time evolution (ITE) on a matrix product state (MPS), incorporating non-local couplings through structured SWAP networks and spectral qubit mapping of logical qubits. The SWAP networks, composed exclusively of local two-qubit gates, effectively mediate non-local qubit interactions. We investigate two distinct network architectures based on rectangular and triangular meshes of SWAP gates and analyze their performance in combination with spectral qubit ordering, which maps logical qubits to MPS sites based on the Laplacian of the logical qubit connectivity graph. The proposed framework is evaluated on synthetic MaxCut instances with varying graph connectivity, as well as on a dynamic portfolio optimization problem based on real historical asset data involving 180 qubits. On certain problem configurations, we observe an over 20× reduction in error when combining spectral ordering and triangular SWAP networks compared to optimization with shuffled qubit ordering. Furthermore, an analysis of the entanglement entropy during portfolio optimization reveals that spectral qubit ordering not only improves solution quality but also enhances the total and spatially distributed entanglement within the MPS. These findings demonstrate that exploiting problem structure through spectral mapping and efficient routing networks can substantially enhance the performance of tensor-network-based optimization algorithms.

\end{abstract}

\section{Introduction}
Combinatorial optimization problems constitute a class of computationally difficult problems, as the number of possible solutions typically grows exponentially with the number of decision variables. Many real-world tasks, including protein folding~\cite{irback_using_2024, lewis_predicting_2021, romero_protein_2025}, logistics and supply chain optimization~\cite{tambunan_quantum_2023, weinberg_supply_2023}, and portfolio optimization~\cite{lang_strategic_2022, mugel_dynamic_2022}, can be formulated as combinatorial optimization problems. Consequently, even modest improvements in algorithmic performance, such as higher solution quality or reduced runtime, can have a significant impact across diverse fields.

The ongoing development of quantum hardware has motivated the development of numerous algorithms for solving combinatorial optimization problems, with experimental demonstrations on trapped-ion systems~\cite{morris_performant_2024, pagano_quantum_2020, romero_protein_2025}, superconducting qubit architectures~\cite{zhu_combinatorial_2025, weidenfeller_scaling_2022, sack_large-scale_2024}, and quantum annealing devices~\cite{yarkoni_quantum_2022, quinton_quantum_2025}. These algorithms are often quantum–classical hybrid in nature and include, among others, the Quantum Approximate Optimization Algorithm (QAOA)~\cite{farhi_quantum_2014}, the Variational Quantum Eigensolver (VQE)~\cite{peruzzo_variational_2014}, and variational implementations of Quantum Imaginary Time Evolution (QITE)~\cite{morris_performant_2024, mcardle_variational_2019}. While these methods have demonstrated encouraging results on small-scale benchmark instances, extending them to industry-relevant problem sizes remains a major challenge, primarily due to limited qubit counts, gate fidelity, and decoherence on current noisy intermediate-scale quantum (NISQ) hardware~\cite{abbas_challenges_2024}.

In parallel with advances in quantum computing, the development of improved classical simulation methods for quantum systems has also accelerated. Tensor network methods have proven especially effective for simulating quantum systems with limited entanglement~\cite{vidal_efficient_2003, orus_practical_2014}, occasionally challenging claims of quantum advantage in tasks such as Hamiltonian simulation~\cite{tindall_efficient_2024} and Gaussian boson sampling~\cite{oh_classical_2024}. Among these methods, Matrix Product States (MPS) are among the most widely studied tensor network architectures, consisting of a one-dimensional chain of tensors connected through virtual bonds that encode correlations~\cite{orus_practical_2014, vidal_efficient_2003, vidal_efficient_2004}. Their structure enables efficient normalization and application of local gate operations through canonical forms, making them particularly suitable for simulating systems with local interactions. More recently, quantum-inspired optimization techniques based on tensor networks, and in particular on matrix product states, have been proposed for solving combinatorial optimization problems~\cite{sreedhar_quantum_2022, feeney_mps-juliqaoa_2025}. However, the inherently one-dimensional topology of MPS introduces limitations, as simulating systems with long-range logical-qubit couplings or correlations remains challenging.

In this study, we investigate methods for addressing optimization problems by employing imaginary time evolution with an MPS state representation. In particular, we evolve the state in imaginary time by using SWAP networks, which allow us to efficiently simulate all-to-all logical qubit connectivity using only nearest-neighbor two-qubit gates. We explore different SWAP network architectures based on rectangular and triangular meshes of SWAP gates paired with a strategy for improved logical qubit mapping to MPS sites based on spectral ordering. To assess the efficacy of the proposed method, we consider a variety of increasingly complex combinatorial optimization problems. First, we evaluate the MaxCut problem on 3-regular (3Reg),  Erdős-Rényi (ER), and fully-connected Sherrington-Kirkpatrick (SK) graphs in order to assess performance on optimization problems with varying logical qubit connectivity and structure. Furthermore, to assess the effectiveness of the proposed method in a more practical environment, we address a case of dynamic portfolio optimization on a problem instance involving 180 qubits based on real historical asset data.

Our results find substantial performance differences depending on the choice of logical qubit mapping and SWAP network architecture. On certain problem instances, such as for 3Reg graphs, we find an over 20× reduction in the error when employing a logical qubit mapping based on spectral ordering, as compared to a shuffled mapping. Interestingly, the benefit of using an optimized logical qubit mapping generally appears to be more pronounced when paired with  a triangular SWAP network. These results indicate that one must consider logical qubit mapping to MPS sites in conjunction with SWAP network architecture to achieve optimum performance. On the dynamic portfolio optimization problem, we acquire an error of merely $0.40\%$ relative to the state-of-the-art Gurobi optimizer, indicating that the proposed method generalizes well beyond synthetic MaxCut instances. Interestingly, we also find that spectral ordering of logical qubits not only improves solution quality, but also results in significantly larger amounts of entanglement being introduced into the MPS during optimization.

In summary, the main contributions of our work are:
\begin{itemize}
    \item We demonstrate optimization on problem instances with complex logical qubit connectivity graphs using SWAP networks to handle long-range interactions, examining the impact of different SWAP network architectures in conjunction with improved logical qubit mapping to MPS sites. To the best of our knowledge, this is the first study that utilizes a triangular SWAP network alongside an MPS representation for quantum simulation of non-local Hamiltonians.
    \item We investigate performance benefits of using problem-specific logical qubit mapping to MPS sites based on the Laplacian of the qubit connectivity graph, occasionally finding an over 20× reduction in the error compared to shuffled logical qubit mapping.
    \item We evaluate our method on MaxCut instances with diverse qubit connectivity and on a portfolio optimization problem from computational finance, demonstrating both robust performance and strong generalization beyond synthetic MaxCut benchmarks.
    \item We examine the correlation between solution quality and entanglement entropy over bipartitions of the MPS during optimization, finding that improved logical qubit mapping increases both the total exhibited entanglement and the spatial distribution of entanglement over the bonds of the MPS.
\end{itemize}

\section{Preliminaries}
\begin{figure}[H]
  \centering
  \includegraphics[width=\columnwidth]{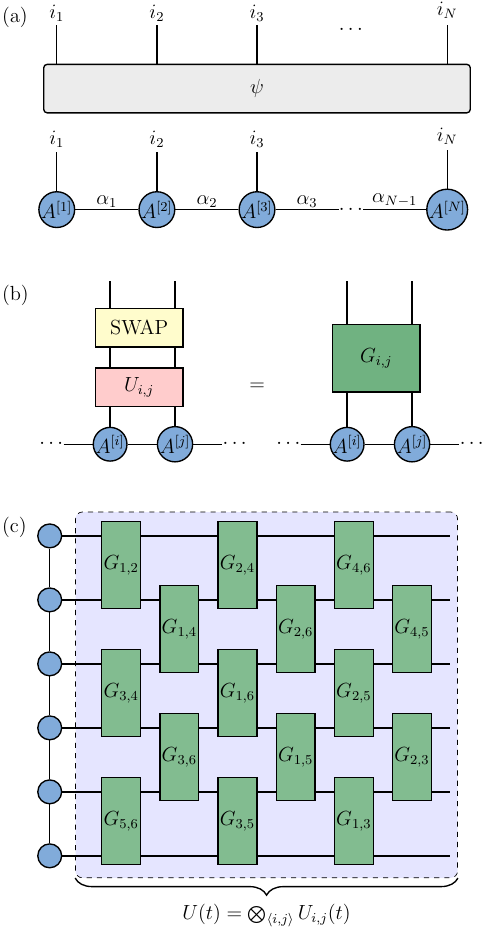}
  \caption{\textbf{(a)} An $N$-qubit quantum state for which the probability amplitudes are encoded in a rank-$N$ tensor $\psi_{i_1...i_N}$ can be decomposed as an MPS, representing a product of lower-rank tensors contracted in a 1D-chain. \textbf{(b--c)} A time evolution operator of commuting long-range interactions $U(t) = \bigotimes_{\langle i, j\rangle}U_{i,j}(t)$ can be exactly factorized as a mesh of local interactions, as shown in (c). Each operator $G_{i,j}$ in the grid applies an interaction $U_{i,j}$ between a logical qubit pair $(i,j)$ followed by a SWAP-gate, as depicted in (b). This way, the order of the logical qubits on the sites of the MPS is permuted in each layer of the mesh.}
  \label{fig:MPS & SWAP illustration}
\end{figure}
In the following section, we formally introduce the notion of \textit{matrix product state} (MPS) in Section~\ref{sec: MPS} and optimization via imaginary time evolution in Section~\ref{sec: QITE}.  Furthermore, we discuss the factorization of time evolution operators using SWAP networks in Section~\ref{sec: SWAP Network}. Finally, we provide an overview of related work in Section~\ref{sec: related work}.
\subsection{Matrix product states (MPS)}
\label{sec: MPS}
An arbitrary $N$-qubit state $\ket{\psi}$ can be expressed in the computational basis as
\begin{equation}
    \ket{\psi} = \sum_{i_1, i_2, ..., i_{N}=0}^1 \psi_{i_1 i_1 ...i_N}\ket{i_1 i_2 ...i_{N}}
\end{equation}
where the complex-valued probability amplitudes of the basis states are encoded in the rank-$N$ tensor $\psi_{i_1i_2...i_N}$. By performing a sequence of singular-value decompositions~\cite{vidal_efficient_2003}, the tensor $\psi_{i_1i_2...i_N}$ can be factorized as a product of lower-rank tensors
\begin{equation}
    \psi_{i_1i_2...i_N} = \sum_{\alpha_1, \alpha_2,..., \alpha_N-1}^{\chi}A^{[1]}_{i_1 \alpha_1} A^{[2]}_{\alpha_1 i_2 \alpha_2}...A^{[N]}_{\alpha_{N-1} i_{N}}.
    \label{eq:MPS}
\end{equation}
A tensor factorization in the form of~(\ref{eq:MPS}) is known as a \textit{matrix product state} (MPS), which is composed of rank-3 tensors arranged along a one-dimensional chain, as depicted in Fig.~\ref{fig:MPS & SWAP illustration} (a). We refer to $\{i_1, i_2, ..., i_{N}\}$ as \textit{physical indices} and to $\{\alpha_1, \alpha_2, ..., \alpha_{N-1}\}$ as \textit{virtual indices}. The dimension of the virtual indices is referred to as the \textit{bond dimension} $\chi$. Although the factorization in~(\ref{eq:MPS}) always provides an exact representation of the original tensor $\psi_{i_1 i_1 ...i_N}$ if we let $\chi$ be exponentially large in the system size~\cite{orus_practical_2014}, an approximate representation of the state can be achieved by truncating the bond dimension $\chi$ to retain only the largest singular values. Crucially, such a truncated MPS can be represented using just $\mathcal{O}(N\chi^2)$ parameters, thereby circumventing the need for an exponential number of parameters required to represent the state exactly. 

\subsection{Imaginary time evolution (ITE)}
\label{sec: QITE}
The evolution of a quantum state $\ket{\psi(\tau)}$ over an imaginary time interval $\Delta \tau$ generated by a Hamiltonian $H$ is expressed as
\begin{equation}
\ket{\psi(\tau_0 + \Delta\tau)} = \frac{e^{-\Delta \tau H}\ket{\psi(\tau_0)}}{\sqrt{\braket{\psi(\tau_0)|e^{-2\Delta\tau H}|\psi(\tau_0)}}}.
\label{eq:exact ITE evolution}
\end{equation}
where we let $\ket{\psi(\tau_0)}$ denote the initial state. The normalization in~(\ref{eq:exact ITE evolution}) is necessary due to the non-unitarity of the imaginary time evolution operator $\exp(-\Delta\tau H)$. Physically, evolving in imaginary time exponentially suppresses the amplitudes of excited states, causing the state to converge to the ground state:
\begin{equation}
\lim_{\tau \to \infty} \ket{\psi(\tau)} = \ket{\psi_\text{gs}},
\end{equation}
provided the initial state has finite overlap with the ground state, $|\braket{\psi(\tau_0)|\psi_\text{gs}}| \neq 0$~\cite{carr_understanding_2010}. 

For combinatorial optimization problems, where the cost is encoded in a Hamiltonian whose ground state represents the optimal solution, imaginary time evolution provides an exact procedure to reach the optimum. While imaginary time evolution is challenging to implement on real quantum hardware due to its non-unitary nature, it is particularly well-suited for tensor network simulations, where operations need not be unitary. Nevertheless, the evolution can generate substantial entanglement~\cite{morris_performant_2024}, making efficient representation of the state along the imaginary time trajectory computationally demanding.

\subsection{SWAP networks}
\label{sec: SWAP Network}
We define a \textit{SWAP network} as a quantum circuit consisting of SWAP gates acting on a set of qubits such that every pair of logical qubits becomes adjacent at some point in the network~\cite{kivlichan_quantum_2018}. As we are working with an MPS state representation, two logical qubits are considered adjacent when they are mapped to adjacent MPS sites. Although we consider only two-qubit interactions, generalizations of SWAP networks to accommodate higher-order terms have also been studied~\cite{ogorman_generalized_2019}. 

Crucially, for a time-evolution operator $U(t)$ that factorizes into commuting two-qubit terms,
\begin{equation}
    U(t) = \bigotimes_{\langle i, j\rangle} U_{i,j}(t),
\end{equation}
the evolution can be implemented by applying each gate $U_{i,j}(t)$ at a point in the SWAP network where the logical qubit pair $(i,j)$ is adjacent, as illustrated in Fig.~\ref{fig:MPS & SWAP illustration} (b--c). The power of a SWAP network therefore lies in its ability to simulate effective all-to-all logical qubit connectivity using only nearest-neighbor interactions, with circuit depth scaling linearly as $\mathcal{O}(N)$. This makes them useful both for quantum hardware with limited physical qubit connectivity and, in our case, for MPS-based simulations, where applying non-local two-qubit gates can be done efficiently despite the one-dimensional topology of the MPS.

\begin{figure*}[ht]
  \centering
  \includegraphics[width=\textwidth]{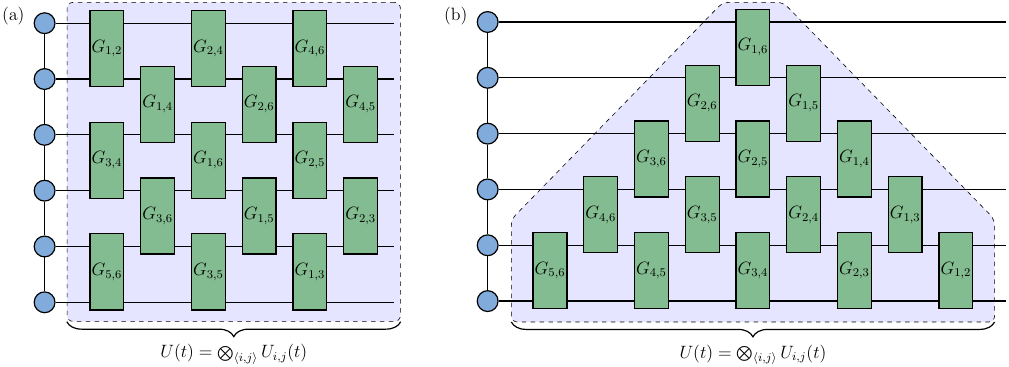}
  \caption{\textbf{(a)} Rectangular and \textbf{(b)} triangular SWAP networks used to factorize an imaginary time evolution operator $U(t)=\exp{(-\Delta\tau H)}$ generated by a non-local Ising Hamiltonian $H = \sum_{\langle i, j\rangle}J_{i,j}Z_iZ_j$. In the depicted circuits, we let $G_{i,j} = \text{SWAP}_{i,j}\circ U_{i,j}$ where $U_{i,j} = \exp{(-\Delta\tau  J_{i,j} Z_i Z_j)}$. The rectangular and triangular decompositions are both exact factorizations of $U(t)$ but permute logical qubits differently within the networks.}
  \label{fig:RSN vs TSN}
\end{figure*}

In this work, we focus on two distinct SWAP network architectures: the \textit{rectangular SWAP network} (RSN) and the \textit{triangular SWAP network} (TSN) (see Fig.~\ref{fig:RSN vs TSN}). These network architectures are inspired by comparable networks of interferometers employed in linear optics to produce arbitrary unitary operations on a collection of photonic modes~\cite{reck_experimental_1994, clements_optimal_2016}. In the RSN, SWAP gates are arranged in a rectangular mesh such that for $N$ qubits, the depth is exactly $N$. In the TSN, the gates follow a triangular pattern, yielding a depth of $2N-3$. In both cases, the network contains precisely $N(N-1)/2$ SWAP gates while maintaining linear depth in $N$.

For exact quantum simulation, the choice of SWAP network is insignificant since both RSN and TSN implement equivalent decompositions of the same unitary operator. However, in approximate MPS-based simulations with a truncated bond dimension, the network structure affects how logical qubits and their correlations are permuted during time evolution, potentially leading to differences in accuracy.

\subsection{Related work}
\label{sec: related work}
In previous work, SWAP networks, originally proposed for simulating electronic structures~\cite{kivlichan_quantum_2018}, have been employed to enable MPS-based simulations of QAOA~\cite{sreedhar_quantum_2022}, achieving up to $p=100$ QAOA layers on Erdős–Rényi graphs with up to 60 vertices. In this work, we instead employ imaginary time evolution as the optimization routine, which naturally drives the state toward the ground state of the cost Hamiltonian and circumvents the challenge of selecting suitable variational parameters $(\gamma, \beta)$ required in QAOA-based optimization.

In terms of algorithmic development, several methods have been proposed to accelerate the convergence of imaginary-time tensor network evolution. Examples include techniques based on environment recycling, which reuse local environments near convergence to reduce computational overhead~\cite{phien_fast_2015}, and more recent approaches introducing reflection “boosts” to accelerate descent toward lower-energy states~\cite{symons_boosted_2024}. For combinatorial optimization including constraints, progress has also been made in designing tensor networks that are intrinsically forced to remain in the feasible subspace, enabling optimization without the need for penalty-based constraint encoding~\cite{lopez_piqueres_cons-training_2025, nakada_quick_2025}. While our focus is on studying the effects of logical qubit mapping and routing on the performance of MPS-based imaginary time evolution, these algorithmic acceleration and constraint-handling techniques could be incorporated into our framework in future work.

On the application side, tensor network methods combined with imaginary time evolution have been proposed as powerful solvers for combinatorial optimization problems with local constraints on the decision variables~\cite{hao_quantum-inspired_2022}.  In another study, MPS-based imaginary-time evolution was applied to the problem of dynamic portfolio optimization~\cite{mugel_dynamic_2022}, solving binary formulations with more than 1000 decision variables and achieving results comparable to, or surpassing, both classical methods and quantum annealing solutions. We similarly adopt a portfolio optimization problem as a benchmark to validate our approach in a realistic setting, ensuring that the observed performance trends generalize beyond synthetic problem instances.

\section{Method}
\subsection{Problem formulation}
To evaluate the proposed method, we consider \textit{quadratic unconstrained binary optimization} (QUBO) problems. While many optimization problems can be expressed as QUBOs, it is worth emphasizing that our method can in principle also handle integer-valued decision variables by extending the dimension of the physical indices of the MPS. 

Formally, a QUBO is defined as a minimization problem
\begin{equation}
    \min_\mathbf{x} \mathbf{x}^TQ\mathbf{x} = \min_{\mathbf{x}} \sum_{i,j}x_i Q_{i,j}x_j,
\end{equation}
where $\mathbf{x}\in \{0,1\}^N$ is a binary vector and $Q\in \mathbb{R}^{N\times N}$ is an upper triangular matrix with real entries~\cite{glover_quantum_2022}.  It is often convenient to rewrite the problem in terms of spin variables $z_i \in \{\pm 1\}$, obtained via the transformation $z_i = 1-2x_i$~\cite{lucas_ising_2014}. This gives an equivalent Ising formulation of the cost function 
\begin{equation}
    \min_{\mathbf{z}} \sum_{\langle i,j \rangle} J_{i,j} z_i z_j + \sum_{i} h_i z_i,
    \label{eq:Ising_formulation}
\end{equation}
where we let $\langle i,j\rangle$ denote the set of coupled logical qubit pairs. The Ising formulation in~(\ref{eq:Ising_formulation}) maps directly to a cost Hamiltonian by replacing binary variables with Pauli-$Z$ operators:
\begin{equation}
    H = \sum_{\langle i, j \rangle} J_{i,j} Z_i Z_j + \sum_i h_i Z_i.
    \label{eq:Ising cost hamiltonian}
\end{equation}
The optimization task then corresponds to finding the ground state of $H$.  

\subsubsection{MaxCut}
As an initial benchmark, we consider the MaxCut problem. The Ising Hamiltonian of the MaxCut problem takes the form
\begin{equation}
    H_\text{MaxCut} = -\sum_{\langle i, j \rangle} \frac{W_{i,j}}{2}\left(1-Z_i Z_j\right),
\end{equation}
where $W_{i,j}$ is the edge weight corresponding to a vertex pair $(i,j)$ and the minus sign ensures a minimization objective. We study MaxCut on three categories of graphs:
\begin{itemize}
    \item 3-regular (3Reg) graphs with unit edge weights,
    \item Erdős–Rényi (ER) graphs with unit edge weights and edge probability $p=0.5$,
    \item fully connected Sherrington-Kirkpatrick (SK) graphs with edge weights $\pm1$ chosen uniformly at random.
\end{itemize}
This selection of graphs provides us with a robust benchmark on graphs with varying qubit connectivity.

\subsubsection{Portfolio optimization}
In order to evaluate performance in a more realistic setting, we consider a dynamic portfolio optimization problem including transaction costs. The problem is formulated as follows.
Given a set of assets \(n \in \{1,2,\ldots,N_a\}\) and discrete time steps \(t \in \{1,2,\ldots,N_t\}\), 
the objective is to optimize the amount \(\omega_{n,t}\) invested in every asset \(n\) at every time \(t\). 
Because the decision variables are encoded in qubits, we express each position in a binary representation normalized by the total available funds \(K\),
\begin{equation}
    \omega_{n,t} = \frac{1}{K}\sum_{q=1}^{N_q} 2^{q-1} x_{n,t,q}.
    \label{eq: binary invested amount}
\end{equation}
In~(\ref{eq: binary invested amount})  $N_q$ is the number of binary variables used to represent $\omega_{n, t}$ and each binary variable $x_{n,t,q} \in \{0,1\}$ is represented by a single qubit. This way, $\omega_{n,t}$ is interpreted as the fraction of the available funds $K$ invested in a given asset $n$ at time $t$.

The objective is to maximize the risk-adjusted return by minimizing the cost function
\begin{equation}
\begin{split}
    \sum_{t=1}^{N_t}\!\left(
        -\mu_t^{\top}\omega_t 
        + \gamma\,\omega_t^{\top}\Sigma_t\omega_t
        + \nu|\Delta\omega_t|
    \right),\\
    \text{subject to}\quad
    \sum_n \omega_{n,t} = 1, \quad \forall t,
\end{split}
\label{eq: portfolio optimization exact}
\end{equation}
where $\mu_t$ is the vector of logarithmic returns, and $\Sigma_t$ is the covariance matrix of the considered assets at time $t$. We refer to $\gamma$ as the risk-aversion parameter, whereas $\nu$ encodes the proportional transaction cost. The budget constraint encodes the fact that we desire the full amount of available funds $K$ to be invested at any given time, which can be embedded into the QUBO formulation using a quadratic penalty term. Furthermore, we approximate the transaction cost with a parabolic term,
\begin{equation}
    \nu|\Delta\omega_t| \approx \zeta (\Delta\omega_t)^2,
\end{equation}
where the coefficient $\zeta$ is chosen such that the quadratic function best approximates the linear cost over the admissible range of trade volumes $[0, K'/K]$, with $K' = 2^{N_q} - 1$. We acquire such an approximation by minimizing the integral
\begin{equation}
    \int_0^{K'/K}\left(\nu x - \zeta x^2\right)^2dx \implies \zeta=  \frac{5}{4}\nu\frac{K}{K'}.
\end{equation}
The complete cost function in a QUBO formulation thus becomes
\begin{equation}
\begin{split}
&\sum_{t=1}^{N_t}\!\left(
-\mu_t^{\top}\omega_t 
+ \gamma\,\omega_t^{\top}\Sigma_t\omega_t
+ \rho\left(u^{\top}\omega_{t}-1\right)^2 \right)
\\
{}+ &\sum_{t=1}^{N_t-1} \zeta\,(\omega_{t+1}-\omega_{t})^2
\end{split}
\label{eq:portfolio QUBO}
\end{equation}
where $u$ is defined as a vector of length $N_a$ whose entries are all ones. In~(\ref{eq:portfolio QUBO}) we do not explicitly write the summation over the assets to simplify the notation.

\subsection{Imaginary time evolution using SWAP networks}
As the optimization routine, we employ imaginary time evolution (ITE) implemented using SWAP networks. The ITE operator $\exp(-\Delta\tau H)$ generated by the Hamiltonian in (\ref{eq:Ising cost hamiltonian}) can be exactly factorized into a product of single- and two-qubit non-unitary gates,
\begin{equation}
    e^{-\Delta\tau H}
    = \bigotimes_{i=1}^N e^{-\Delta\tau h_i Z_i}
      \bigotimes_{\langle i, j \rangle} e^{-\Delta\tau J_{i,j} Z_i Z_j},
    \label{eq: factorize_evol_operator}
\end{equation}
without requiring a Trotter-Suzuki decomposition, as all terms in the Hamiltonian mutually commute.
Within the SWAP network, each two-qubit gate $\exp(-\Delta\tau J_{i,j} Z_i Z_j)$ is contracted onto the MPS when logical qubits \( i \) and \( j \) are adjacent. The order in which gates are contracted is illustrated for both rectangular and triangular SWAP networks in Fig.~\ref{fig:RSN vs TSN}. Before contracting a two-qubit gate onto the MPS, the orthogonality center is shifted to one of the qubits on which the gate acts in order to easily restore the canonical form of the MPS after the gate contraction. Further details on the canonical form of the MPS are provided in Appendix~\ref{appendix: canonical form}.

Two-qubit gates are applied to the state by contracting the gate with the physical indices of the corresponding MPS sites. The contraction is followed by a singular value decomposition (SVD) and truncation of the resulting bond dimension such that only the $\chi$ largest singular values are kept (see Appendix~\ref{appendxi: 2-qubit gate} for details). This ensures that the MPS remains computationally efficient while approximating the true imaginary time trajectory. If there is no interaction term present between a given pair of logical qubits, only a SWAP gate is applied. If the cost Hamiltonian includes linear terms, the corresponding single-qubit gates are applied after the SWAP network. It is worth noting that the order of the logical qubits will be reversed after all gates within the SWAP network have been applied.

\subsection{Qubit mapping via spectral ordering}
Before optimization can proceed, logical qubits must be mapped to the sites of the MPS. Since the MPS has a one-dimensional geometry in which correlations decay exponentially with separation distance~\cite{orus_practical_2014}, it is desirable to place strongly coupled qubits that are likely to share entanglement on neighboring sites. However, acquiring such a mapping is often a challenging task, as the interaction graph of the qubits may be complex and difficult to accurately embed in a one-dimensional geometry.

A natural formulation of the qubit-mapping problem is to minimize the cost function
\begin{equation}
    C\left(\{\pi_i\}_{i=1}^N\right) = \sum_{\langle i, j \rangle} A_{i,j} (\pi_i - \pi_j)^2,
    \label{eq: integer mapping cost}
\end{equation}
where $A_{i,j}$ denotes the interaction strength between a logical qubit pair $(i,j)$, and $\pi_i \in \{1,\dots,N\}$ denotes the assigned MPS site of logical qubit $i$. This cost penalizes strongly coupled qubits that are placed far apart and should therefore yield an approximate one-dimensional embedding. However, due to the discrete nature of the variables, minimizing \eqref{eq: integer mapping cost} is NP-hard~\cite{george_analysis_1997}, thus making it impractical for large system sizes.

A useful relaxation of the problem arises from spectral graph theory. Let $L = D - A$ be the graph Laplacian, where $A$ is the weighted adjacency matrix and $D$ is the diagonal degree matrix. The continuous optimization problem
\begin{equation}
    \begin{split}
    \text{minimize} \quad & h(\mathbf{x}) = \mathbf{x}^T L \mathbf{x}
    = \sum_{\langle i,j \rangle} A_{i,j} (x_i - x_j)^2 \\
    \text{subject to} \quad & \sum_i x_i = 0, \quad \sum_i x_i^2 = 1,
    \end{split}
    \label{eq: fiedler mapping cost}
\end{equation}
is solved by the eigenvector of $L$ corresponding to its second-smallest eigenvalue, known as the \emph{Fiedler vector}~\cite{atkins_spectral_1998}. The constraints in~(\ref{eq: fiedler mapping cost}) fix normalization and eliminate the trivial solution. 

Since~(\ref{eq: fiedler mapping cost}) can be seen as a continuous relaxation of \eqref{eq: integer mapping cost}, the Fiedler vector provides a heuristic ordering of the logical qubits. We sort the entries of the Fiedler vector and assign qubits to MPS sites according to this order. For the interaction strength in~(\ref{eq: fiedler mapping cost}), we consider the magnitude of the Ising coupling $A_{i,j} = |J_{i,j}|$. This \emph{spectral ordering} of the logical qubits serves as an efficient pre-processing step, requiring only an eigenvalue decomposition of an $N\times N$ matrix.

\subsection{Increasing numerical stability}
\begin{figure*}[ht]
  \centering
  \includegraphics[width=\linewidth]{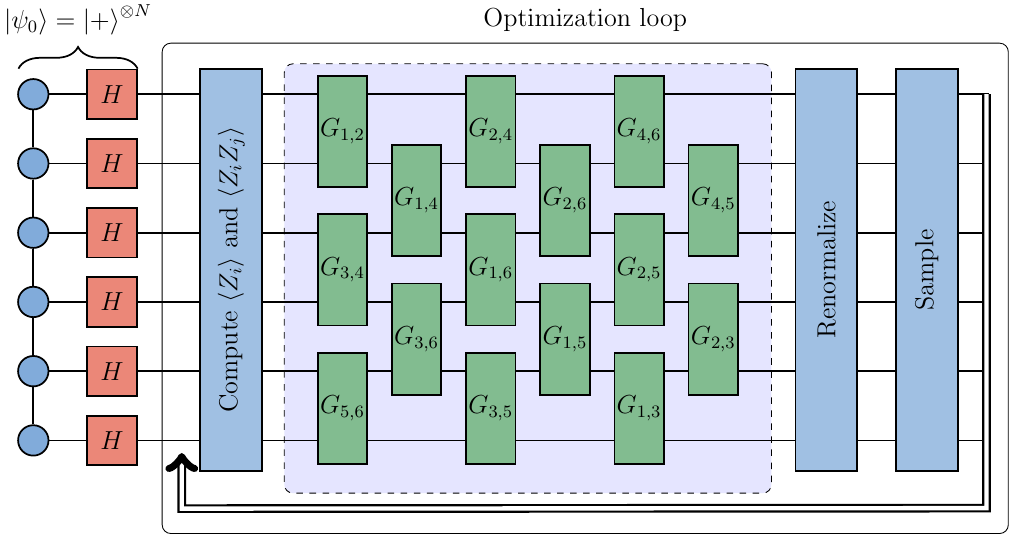}
  \caption{A depiction of the optimization loop. The initial state $\ket{\psi_0} = \ket{+}^{\otimes N}$ state is prepared by initializing the MPS in the $\ket{0}^{\otimes N}$ state followed by a contraction of a Hadamard gate on each physical index. The optimization loop recursively updates the state as $\ket{\psi_{s+1}}= \exp(-\Delta\tau (H-\braket{H}_s)\ket{\psi_s}$ with renormalization and sampling following each state update. The lowest energy sample of each iteration is kept as the output.}
  \label{fig:optimization loop}
\end{figure*}

Due to the non-unitarity of the imaginary time evolution operator \(\exp(-\Delta \tau H)\), its direct contraction on an MPS can cause the tensor entries to either grow or decay exponentially, leading to numerical instabilities. Renormalizing the MPS after each application of \(\exp(-\Delta \tau H)\) alleviates this issue to some extent, but even a single application can still be unstable, depending on the Hamiltonian and the chosen time step \(\Delta \tau\).  

The dynamics governing the normalized state evolution in ~(\ref{eq:exact ITE evolution}) can be expressed as a non-linear differential equation
\begin{equation}
    \frac{d\ket{\psi}}{d\tau} = -(H-\braket{\psi|H|\psi})\ket{\psi}
    \label{eq:exact QITE differential eq}
\end{equation}
where the subtraction of the expectation value in~(\ref{eq:exact QITE differential eq}) ensures that the state remains normalized throughout time evolution.

To enhance numerical stability during ITE, we employ a discrete-time approximation of the renormalization in~(\ref{eq:exact QITE differential eq}). To do so, we compute the single-site expectation values  
\[
    \langle Z_i \rangle_s = \langle \psi_s | Z_i | \psi_s \rangle ,
\]
and two-site correlators
\[
    \langle Z_i Z_j \rangle_s = \langle \psi_s | Z_i Z_j | \psi_s \rangle .
\]
of the MPS sites. Here, we let $\ket{\psi_s}$ denote the state of the MPS at optimization step $s$. These expectation values can be acquired via exact tensor network contractions in $\mathcal{O}(N\chi^3)$ and $\mathcal{O}(N^2\chi^3)$ time~\cite{fishman_itensor_2022}, respectively. We then update the quantum state using the normalized imaginary time-evolution operator  
\begin{equation}
    |\psi_{s+1}\rangle = \exp\!\Big(-\Delta\tau \sum_{\langle i,j \rangle} J_{i,j}\big(Z_iZ_j - \langle Z_iZ_j \rangle_s\big)\Big)\, |\psi_s\rangle ,
    \label{eq: normalized discrete evolution}
\end{equation}
where each Hamiltonian term is shifted by its expectation value at the current step. If single-qubit terms are present in the Hamiltonian, they are similarly normalized by subtracting $\braket{Z_i}_s$. This normalization of the gates suppresses exponential growth or decay arising from large eigenvalue contributions, thereby mitigating numerical instabilities.  

Since this procedure is only a discrete-time approximation of the continuous-time differential equation~(\ref{eq:exact QITE differential eq}), the state does not remain exactly normalized during time evolution. An explicit renormalization after each update step is therefore still required to ensure accurate sampling and expectation value calculations.

\subsection{Complete algorithm}
The complete optimization routine proceeds as follows. We begin by initializing the quantum state as a balanced superposition of all computational basis states,  
\begin{equation}
    \ket{\psi_0} = \frac{1}{\sqrt{2^{N}}}\sum_{i_1 , \dots, i_{N}=0}^{1}\ket{i_1 \dots i_{N}} ,
    \label{eq:balanced superposition}
\end{equation}
which is a product state and can therefore be exactly represented by an MPS with bond dimension $\chi = 1$.  

The quantum state is then evolved recursively using the following procedure: 
\begin{enumerate}
    \item We evaluate single-site expectation values $\langle Z_i \rangle$ and two-site correlators $\langle Z_i Z_j \rangle$, from which we obtain the expectation values of all terms in the cost Hamiltonian.  
    \item To evolve the quantum state over an interval $\Delta \tau$, we apply the imaginary-time evolution operator using a SWAP network, incorporating gate normalization as presented in~(\ref{eq: normalized discrete evolution}).  
    \item  We renormalize the MPS representation of the state, with the norm determined by contracting the MPS with itself.
    \item We sample the evolved state $N_s$ times using the \textit{perfect sampling algorithm}~\cite{ferris_perfect_2012}, which has a time complexity of $\mathcal{O}(N_s\chi^2)$. We compute the cost corresponding to each sample and keep the lowest cost sample as the output of the current step.
\end{enumerate}
The optimization procedure described above is depicted graphically in Fig.~\ref{fig:optimization loop}.

As a convergence criterion, we monitor the variance of the sampled energies, $\sigma^2_s$, at each optimization step $s$. The optimization is terminated either when 
\begin{equation*}
    \sigma^2_s < 10^{-3}\,\sigma^2_0,
\end{equation*} 
where $\sigma^2_0$ is the initial energy variance of $\ket{\psi_0}$, or after a preset maximum number of optimization steps $N_{\text{step}}$. Since the imaginary-time Schrödinger equation satisfies  
\begin{equation*}
    \frac{d\braket{H}}{d\tau} \propto -\,\mathrm{Var}(H)\,,
\end{equation*}
using the energy variance of the initial set of samples naturally provides a suitable energy scale for the stopping criterion. Naturally, other threshold values for the stopping criterion can also be used, depending on the desired numerical precision.

The main computational bottleneck of the algorithm arises from executing the SVD during two-qubit gate contractions, which has a time complexity of $\mathcal{O}(\chi^3)$. As the SWAP networks consist of $N(N-1)/2$ two-qubit gates that are contracted in sequence, we anticipate the overall time complexity of the proposed method to be $\mathcal{O}(N^2\chi^3)$. We summarize the full optimization procedure in Algorithm~\ref{alg:MPS-QITE}.
\begin{algorithm}[H]
\caption{MPS-QITE Optimization}
\label{alg:MPS-QITE}
\begin{algorithmic}[1]
\State Initialize quantum state $\ket{\psi_0}$ as balanced superposition~(\ref{eq:balanced superposition})
\State Compute variance $\sigma^2_0$ from $N_s$ samples of $\ket{\psi_0}$
\For{$s = 0, 1, \dots, N_\text{step}-1$}
    \State Compute expectation values $\langle Z_i \rangle_s$, $\langle Z_i Z_j \rangle_s$ via exact contractions.
    \State Update state by applying ITE operator factorized using SWAP network:
    $$
        \ket{\psi_{s+1}} = \exp\!\Big(-\Delta\tau \sum_{\langle i,j \rangle} J_{i,j}(Z_{i,j} - \langle Z_{i,j} \rangle_s)\Big) \ket{\psi_s}
    $$
    \State Renormalize evolved MPS $\ket{\psi_{s+1}}$
    \State Sample state $\ket{\psi_{s+1}}$ $N_s$ times using~\cite{ferris_perfect_2012}
    \State Compute variance $\sigma^2_s$ from samples of $\ket{\psi_s}$
    \If{$\sigma^2_s < 10^{-3}\,\sigma^2_0$}
        \State \textbf{break}
    \EndIf
\EndFor
\State \Return lowest energy sample
\end{algorithmic}
\end{algorithm}

\subsection{Experimental details}
\begin{table*}[ht]
\centering
\setlength{\tabcolsep}{8pt}  % widen column spacing
\renewcommand{\arraystretch}{1.15}  % slightly taller rows for readability
\caption{Summary of experimental parameters used for the MaxCut and portfolio optimization problems. 
Symbols are defined as follows: $N$ - number of qubits, $N_{\text{step}}$ - maximum number of imaginary-time steps, 
$N_s$ - number of samples, $\Delta \tau$ - imaginary time step, 
$N_t$ - number of time steps in the portfolio optimization model, $N_a$ - number of assets, 
$N_q$ - number of qubits per asset, $\gamma$ - risk-aversion, 
$\nu$ - transaction cost, $\rho$ - penalty scalar for budget constraint.}
\begin{tabular}{@{}lccccccccccc@{}}
\toprule
\textbf{Problem} & \textbf{Graph type} & $N$ & $N_{\text{step}}$ & $N_s$ & $\Delta \tau$ & $N_t$ & $N_a$ & $N_q$ & $\gamma$ & $\nu$ & $\rho$ \\ 
\midrule
\textbf{MaxCut} & 3Reg & 100 & 30 & 1000 & 1.0 & — & — & — & — & — & — \\
 & ER  & 100 & 30 & 1000 & 0.06 & — & — & — & — & — & — \\
 & SK  & 100 & 30 & 1000 & 0.03 & — & — & — & — & — & — \\ 
\addlinespace[6pt]
\cmidrule(lr){1-12}
\addlinespace[3pt]
\textbf{Portfolio opt.} & — & 180 & 40 & 1000 & 10 & 9 & 10 & 2 & 1 & 0.01 & 1 \\ 
\bottomrule
\end{tabular}
\label{tab:experimental parameters}
\end{table*}

To evaluate the proposed method on the MaxCut problem, we randomly generate 10 instances each of 3-regular (3Reg), Erdős–Rényi (ER), and Sherrington–Kirkpatrick (SK) graphs with $N=100$ vertices. 

When solving MaxCut on 3Reg and ER graphs, we consider two strategies for initializing the mapping of logical qubits to MPS sites:  
(1) a spectral ordering based on the connectivity graph Laplacian, and  
(2) a randomly shuffled ordering.  
For SK graphs, spectral ordering is not applicable, as the graph is fully connected and all coupling strengths have equal magnitude. Therefore, we only use a shuffled ordering when considering SK graphs.

For the portfolio optimization problem, we construct the QUBO in~(\ref{eq:portfolio QUBO}) based on the actual price history of the top 10 stocks by market capitalization in the S\&P500 index. This data is openly available and summarized in the \textit{quantum optimization benchmarking library}~\cite{koch_quantum_2025}. Besides using spectral and shuffled ordering of the logical qubits, we also test a structured layout where the qubit representing the binary variable $x_{n,t,q}$ is mapped to site
\begin{equation}
\pi_{n,t,q} = q + N_q(n-1) + N_a N_q (t-1),
\label{eq: physically motivated ordering portfolio opt}
\end{equation}
This ordering reflects the structure of~(\ref{eq:portfolio QUBO}), where $\omega_{n,t}$ only couples to assets at the same time $t$ and to the immediately adjacent times $t-1$ and $t+1$. We will refer to this logical qubit mapping as \textit{hierarchical ordering}.

The imaginary time step $\Delta\tau$ is adjusted for each problem instance to ensure efficient and stable convergence. A summary of the time step and other experimental parameters is provided in Table~\ref{tab:experimental parameters}.

We evaluate solution quality using the approximation ratio
\begin{equation}
    \text{AR} = \frac{C_\text{MPS}}{C_\text{ref}},
\end{equation}
where $C_\text{MPS}$ is the cost returned by our MPS-based solver and $C_\text{ref}$ denotes the reference cost. We also report the error $\epsilon$, here defined as simply
\begin{equation}
    \epsilon = 1 - \text{AR}.
\end{equation}
When the provably optimum cost $C_\text{opt}$ is known, we let $C_\text{ref} = C_\text{opt}$; otherwise, we use the best solution obtained from the high-performing BiqCrunch~\cite{krislock_biqcrunch_2017} MaxCut solver as our reference. Additionally, we benchmark all MaxCut instances against the state-of-the-art Gurobi optimization software, confirming that Gurobi and BiqCrunch yield the same costs for all considered problem instances. For 3Reg graphs, we compare against provably optimal solutions. For ER and SK instances, the reference solutions always have an optimality gap of $\leq 1\%$ and $< 10\%$, respectively. On the portfolio optimization problem, we use the solution provided by Gurobi as our reference, with a $9.4\%$ optimality gap.

All MPS simulations are implemented using the \texttt{ITensor} library~\cite{fishman_itensor_2022} and executed on a single CPU thread.

\begin{figure}[H]
  \centering
  \includegraphics[width=\columnwidth]{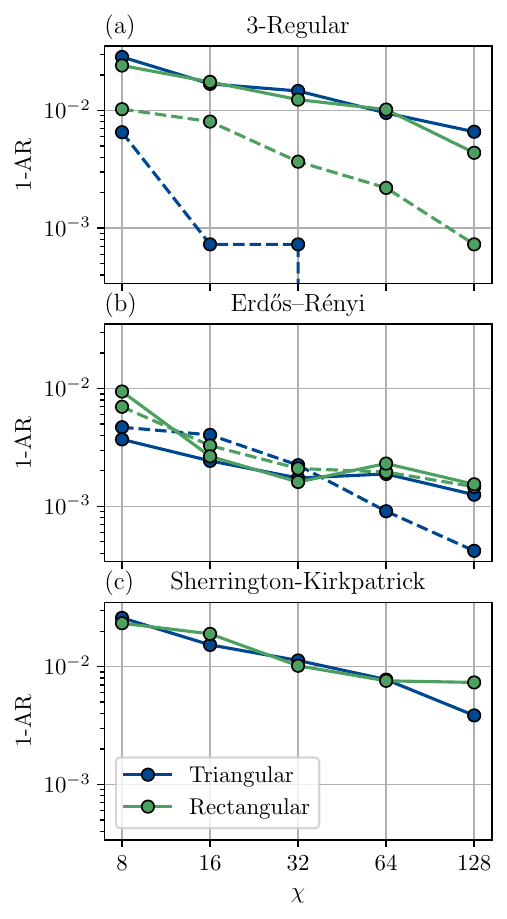}
  \caption{The error $\epsilon = 1 - \mathrm{AR}$ when solving MaxCut on \textbf{(a)} 3-regular, \textbf{(b)} Erdős–Rényi, and \textbf{(c)} Sherrington–Kirkpatrick graphs as a function of the bond dimension $\chi$. Each curve shows the mean error over all considered instances. Solid lines correspond to shuffled mappings of logical qubits to MPS sites, while dashed lines correspond to spectral ordering.}
  \label{fig:AR vs chi}
\end{figure}
\section{Results}
\subsection{MaxCut}
Concerning the numerical results for the MaxCut problem, we first examine how the mean error depends on the bond dimension $\chi$, the SWAP network architecture, and the logical qubit ordering (see Fig.~\ref{fig:AR vs chi}). As expected, the mean error generally decreases with increasing bond dimension $\chi$. An exception is observed for Erdős–Rényi (ER) graphs, where increasing $\chi$ from 32 to 64 unexpectedly leads to a slight increase in the error when using shuffled qubit mapping.

Regarding the SWAP network architecture, we find little to no difference between the rectangular SWAP network (RSN) and the triangular SWAP network (TSN) when using a shuffled logical qubit ordering, except for a slight improvement with TSN on SK graphs at $\chi = 128$. In contrast, when employing spectral qubit ordering, a clear advantage emerges for the TSN architecture. For 3Reg graphs solved with TSN using $\chi = 16$, we acquire an over 20× reduction of the error when using spectral ordering, from $1.68\%$ to $0.07\%$. With RSN, the spectral ordering still reduces the error (from $1.75\%$ to $0.80\%$), although the improvement is less pronounced. Moreover, for $\chi = 64$ and $\chi = 128$ on 3Reg graphs, the combination of TSN and spectral ordering consistently reaches the reference solution cost across all problem instances, which is not achieved when combining spectral ordering with RSN.

\begin{figure}[H]
  \centering
  \includegraphics[width=\columnwidth]{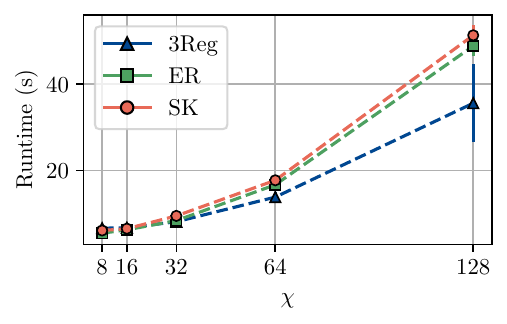}
  \caption{Average runtime (s) per optimization step as a function of bond dimension $\chi$, separated by graph type. Values are reported as mean $\pm$ standard deviation across all network architectures and qubit orderings.}
  \label{fig:runtime}
\end{figure}

For ER graphs, the influence of SWAP network architecture and qubit mapping appears less significant at smaller bond dimensions ($\chi = 8, 16, 32$), as all considered configurations yield very similar results in this regime. However, at larger bond dimensions ($\chi = 64, 128$), the combination of TSN and spectral ordering again provides the lowest mean error, reaching $0.04\%$, while other configurations yield errors in the range of $0.12\%$–$0.15\%$.

\begin{figure}[H]
  \centering
  \includegraphics[width=\columnwidth]{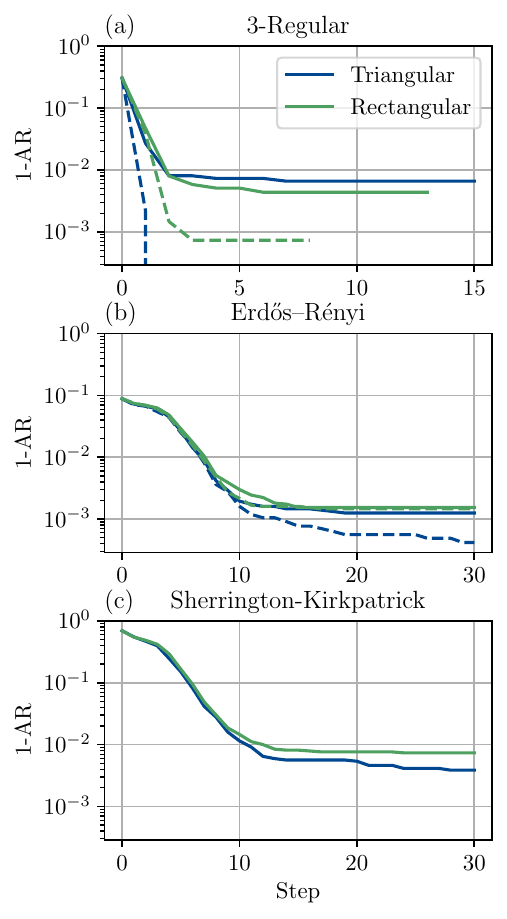}
  \caption{The evolution of the error $\epsilon = 1 - \mathrm{AR}$ during optimization for MaxCut on \textbf{(a)} 3-regular, \textbf{(b)} Erdős–Rényi, and \textbf{(c)} Sherrington–Kirkpatrick graphs. The plotted values represent averages over all problem instances. Solid and dashed lines denote random and spectrally ordered qubit mappings, respectively.}
  \label{fig:AR vs tau}
\end{figure}

\begin{table*}[ht]
\centering
\setlength{\tabcolsep}{6pt}  % adjust column spacing
\renewcommand{\arraystretch}{1.15}  % slightly taller rows
\caption{Mean and standard deviation of the approximation ratio (AR) for MaxCut at bond dimension $\chi = 128$, averaged over 10 graph instances. 
Values are given as mean $\pm$ standard deviation, and numbers in parentheses indicate how many of the 10 instances reached the cost achieved by the reference benchmark.}
\begin{tabular}{@{}llccc@{}}
\toprule
\textbf{Graph} & \textbf{Ordering} & \textbf{3Reg} & \textbf{ER} & \textbf{SK} \\
\midrule
\multirow{2}{*}{Triangular} 
  & Spectral & $1.0000 \pm 0.0000$ (10/10) & $0.9996 \pm 0.0007$ (7/10) & — \\
  & Shuffled & $0.9934 \pm 0.0101$ (6/10) & $0.9987 \pm 0.0016$ (5/10) & $0.9961 \pm 0.0062$ (6/10) \\
\addlinespace[4pt]
\multirow{2}{*}{Rectangular} 
  & Spectral & $0.9993 \pm 0.0023$ (9/10) & $0.9985 \pm 0.0020$ (5/10) & — \\
  & Shuffled & $0.9956 \pm 0.0061$ (6/10) & $0.9985 \pm 0.0027$ (4/10) & $0.9927 \pm 0.0084$ (3/10) \\
\bottomrule
\end{tabular}
\label{tab:maxcut_error}
\end{table*}

A summary of the mean and standard deviation of the AR obtained for $\chi = 128$ is provided in Table~\ref{tab:maxcut_error}. The low standard deviation of the AR, typically ranging from $0.1\%$ to $1\%$, indicates robust performance across different problem instances. The average runtime per optimization step is shown for different values of $\chi$ in Fig.~\ref{fig:runtime}.

It is also informative to examine the rate of convergence of the algorithms. In Fig.~\ref{fig:AR vs tau}, we show the mean error as a function of the optimization step for $\chi = 128$. For 3Reg graphs, we see that using spectral ordering combined with TSN not only produces the best final solutions but also achieves the fastest convergence, requiring at most two optimization steps to match the reference solution. Although the TSN architecture with spectral ordering also achieves the lowest final error for ER graphs, its convergence rate is comparable to that of other configurations. A similar trend is observed for SK graphs: while TSN yields a lower final error than RSN, both architectures exhibit similar rates of convergence.

\subsection{Portfolio optimization}
\begin{figure}[H]
  \centering
  \includegraphics[width=\columnwidth]{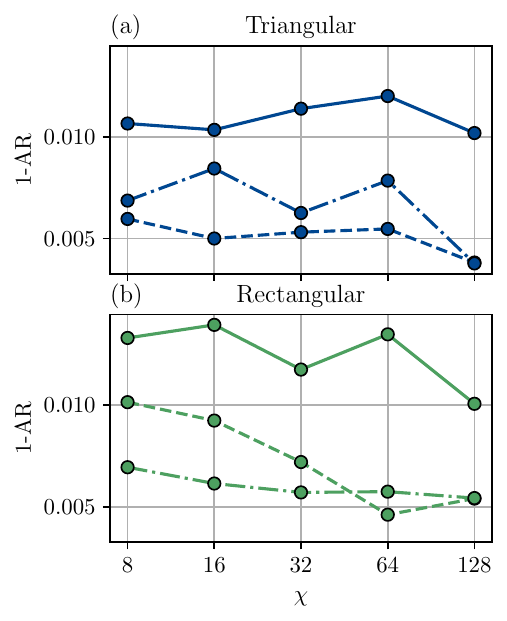}
  \caption{The error $\epsilon = 1 - \mathrm{AR}$ when solving dynamic portfolio optimization as a function of the bond dimension $\chi$ using \textbf{(a)} triangular and \textbf{(b)} rectangular SWAP networks. Solid, dash-dotted, and dashed lines correspond to shuffled, hierarchical, and spectral logical qubit mappings.}
  \label{fig:portfolio AR vs chi}
\end{figure}

For the portfolio optimization problem, we first examine how the error depends on the tensor network architecture, qubit mapping, and bond dimension $\chi$, as summarized in Fig.~\ref{fig:portfolio AR vs chi}. Regarding the qubit mapping, it is evident that the shuffled logical qubit order yields the poorest performance, irrespective of the SWAP network architecture. This result is fairly expected, as a shuffled mapping completely disregards the problem’s intrinsic connectivity structure. Interestingly, the spectral ordering outperforms the hierarchical ordering defined in (~\ref{eq: physically motivated ordering portfolio opt}) when utilizing a TSN architecture and bond dimension $\chi=8,16,32,64$. When  $\chi=128$, the spectral and hierarchical qubit orderings yield an equal error of $0.38\%$. This indicates that the spectral embedding not only improves upon a random qubit mapping but can even surpass heuristic orderings derived from the problem structure. 

When considering an RSN architecture, the spectral ordering performs slightly worse than the hierarchical ordering for bond dimensions $\chi \leq 32$, but yields better or equal performance when $\chi \geq 64$. These findings align well with the results acquired on MaxCut, for which it was found that spectral ordering performs especially well when paired with the TSN architecture. As for the bond dimension, we do not observe a strictly monotonic improvement in performance with increasing $\chi$, although this may be due to the analysis being limited to a single problem instance.

\begin{figure*}[ht]
  \centering
  \includegraphics[width=\textwidth]{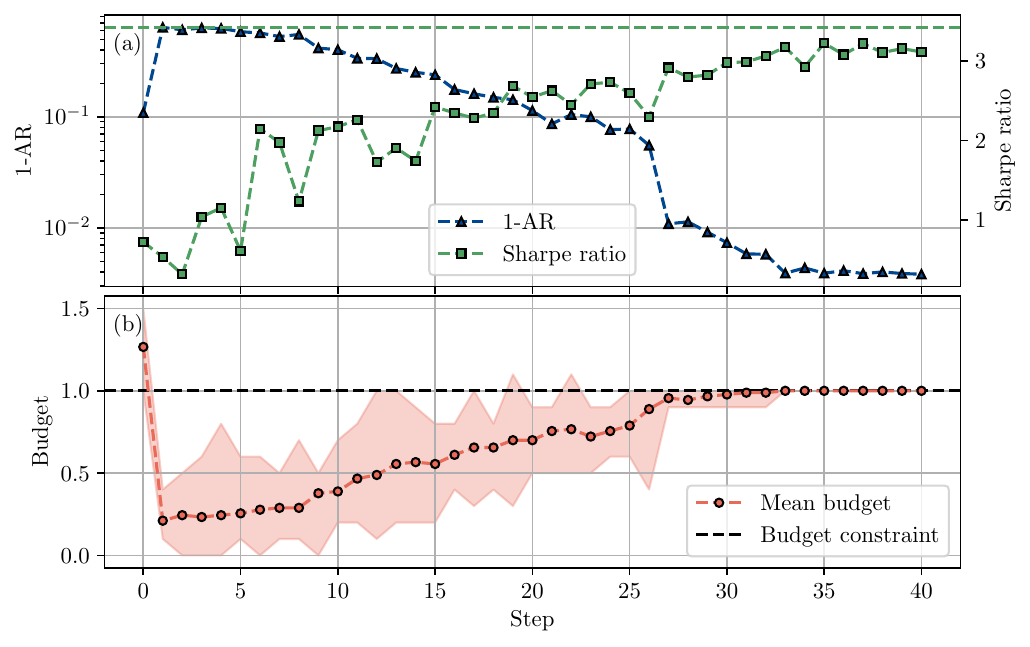}
  \caption{\textbf{(a)} Error and Sharpe ratio as a function of the optimization step. The dashed green line corresponds to the Sharpe ratio of the solution provided by Gurobi. \textbf{(b)} The mean budget of the lowest-cost sample at each optimization step, as defined in~(\ref{eq:average budget}). The shaded area indicates the range between the minimum budget $\min_t \sum_n \omega_{n,t}$ and maximum budget $\max_t \sum_n \omega_{n,t}$ of the lowest cost sample.} 
  \label{fig:portfolio opt performance}
\end{figure*}

Next, we examine the optimization dynamics for the portfolio optimization problem. Specifically, we focus on the configuration employing TSN, spectral qubit ordering, and a bond dimension of $\chi = 128$, as this was previously found to be one of the best-performing setups. As expected, the cost decreases almost monotonically throughout the optimization (see Fig.~\ref{fig:portfolio opt performance} (a)), reaching a final error of $\epsilon = 0.40\%$. This level of accuracy is comparable to, and in some cases even exceeds, that achieved on the synthetic MaxCut instances, indicating that the proposed method generalizes well to more realistic optimization problems.

In addition to the QUBO cost, we also track the Sharpe ratio at each optimization step. The Sharpe ratio is defined as
\begin{equation}
\text{SR} = \frac{\sum_t \mu_t^T \omega_t}{\sqrt{\sum_t \omega_t^T \Sigma_t \omega_t}},
\label{eq:sharpe ratio}
\end{equation}
and is a common performance metric in portfolio optimization, expressing the relative trade-off between expected return and risk. Although the algorithm does not explicitly optimize the Sharpe ratio, we observe a clear increase throughout the optimization, reaching a maximum value of $\text{SR} = 3.22$, similar to the Sharpe ratio of $3.42$ acquired by the high-performing Gurobi solver.

Since the portfolio optimization problem includes an explicit budget constraint, it is also important to verify that this constraint, formulated in~(\ref{eq: portfolio optimization exact}), is satisfied. In Fig.~\ref{fig:portfolio opt performance} (b), we show the mean budget used by the algorithm, defined as
\begin{equation}
\frac{1}{N_t} \sum_{n,t} \omega_{n,t},
\label{eq:average budget}
\end{equation}
at each optimization step. Crucially, we find that the used budget effectively converges to the target throughout optimization. The final solution adheres to the constraint exactly, thus confirming that the algorithm not only reduces the overall cost but also effectively enforces the encoded budget constraint.

\begin{figure*}[ht]
  \centering
  \includegraphics[width=\textwidth]{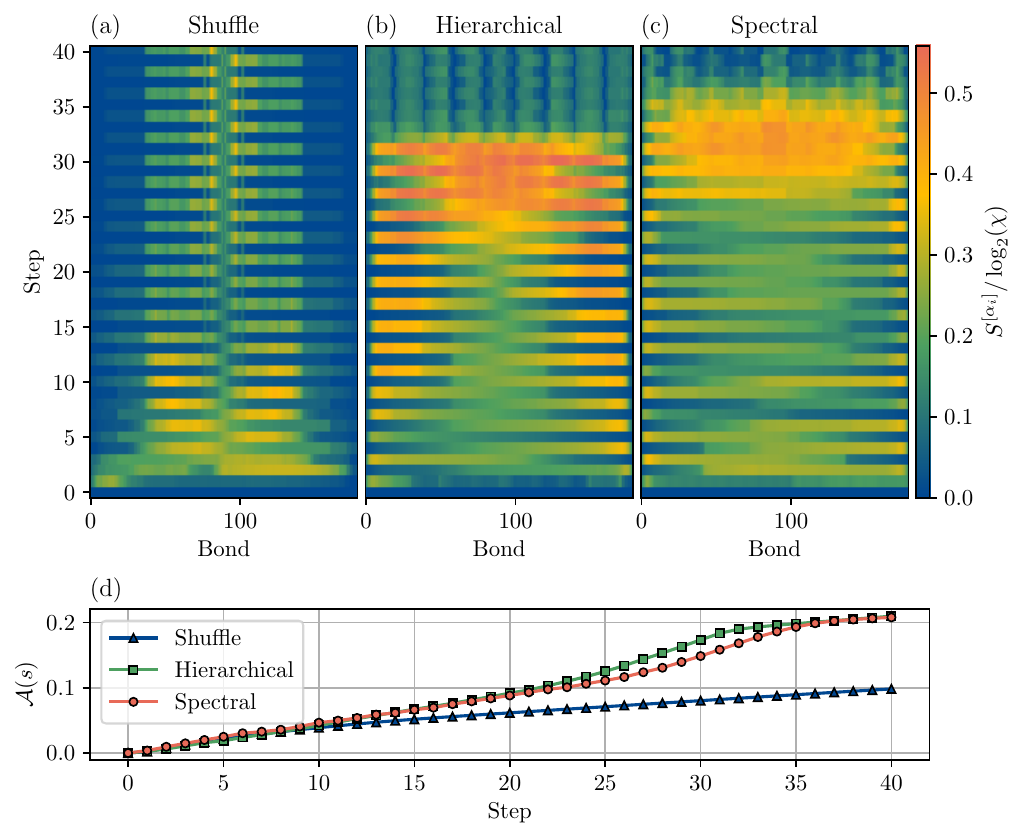}
  \caption{The entanglement entropy calculated at each MPS bond during portfolio optimization using $\chi = 128$ using \textbf{(a)} shuffled, \textbf{(b)} hierarchical, and \textbf{(c)} spectral logical qubit mapping to MPS sites. The spectral and hierarchical mappings, previously found to provide the lowest cost solutions, exhibit substantially larger entanglement entropy more uniformly spread over the MPS bonds compared to the shuffled qubit mapping. The cumulative sum of the entanglement entropy over optimization steps is depicted in \textbf{(d)}, calculated using~(\ref{eq: cumulative entanglement metric}).}
  \label{fig:portfolio entanglement}
\end{figure*}

\subsection{Entanglement study}
One of the advantages of expressing the state as an MPS is that it enables simple computation of the entanglement entropy over bipartitions of MPS sites, in order to quantify the degree of correlation. In Fig.~\ref{fig:portfolio entanglement}, we show the entanglement entropy computed at each bond of the MPS during the portfolio optimization process using a TSN and a bond dimension of $\chi = 128$. The entanglement entropy \(S^{[\alpha_i]}\) across bond \(\alpha_i\) is defined as
\begin{equation}
S^{[\alpha_i]} = -\sum_i s_i^{[\alpha_i]\,2} \log s_i^{[\alpha_i]\,2},
\label{eq:entanglement_entropy}
\end{equation}
where \(s_i^{[\alpha_i]}\) are the Schmidt coefficients obtained from the singular value decomposition of the bipartition \((1,2,\dots,i) | (i+1,\dots,N)\) of the MPS sites.

As seen in Fig.~\ref{fig:portfolio entanglement} (a)--(c), the degree of entanglement exhibited during optimization is substantially higher when using the hierarchical and spectral ordering qubit compared to the shuffled qubit ordering. Optimizing using the hierarchical and spectral ordering not only yields larger peak values of the entanglement entropy but also appears to exhibit a more uniform distribution of entanglement across the bonds, in contrast to the shuffled case where the entropy is concentrated at sites at the center of the MPS. The periodic reflection of the entanglement pattern across the MPS sites between optimization steps can be attributed to the reversal of logical qubit order on the MPS with each application of a SWAP network.

To acquire a metric of the total amount of entanglement exhibited throughout optimization, we consider the cumulative entanglement entropy up to optimization step \(s\), which we define as
\begin{equation}
\mathcal{A}(s) = \frac{\sum_{s'=0}^{s} \sum_{i = 1}^{N-1} S^{[\alpha_i]}_{s'}}{N_\text{step}(N-1)\log_2\chi}.
\label{eq: cumulative entanglement metric}
\end{equation}
Here, we let $S^{[\alpha_i]}_{s'}$ denote the entanglement entropy calculated using~(\ref{eq:entanglement_entropy}) at step $s'$. The cumulative measure $\mathcal{A}(s)$ quantifies the total entanglement produced during the optimization process up to step $s$, normalized by the total number of optimization steps $N_\text{step}$ and the number of bonds $N-1$. Additionally, we normalize by $\log_2 \chi$, which is the largest possible value of the entanglement entropy for a given bond dimension $\chi$. From this cumulative measure, depicted in Fig.~\ref{fig:portfolio entanglement} (d), it is evident that both the hierarchical and spectral orderings accumulate substantially more total entanglement than the shuffled logical qubit ordering throughout optimization.

\section{Discussion}
Concerning the issue of logical qubit mapping, our results highlight the advantage of using spectral ordering as a heuristic for mapping highly connected qubit interaction graphs to MPS sites. For 3Reg graphs, this effect is substantial, sometimes yielding over a 20× reduction in the error compared to shuffled logical qubit mapping. Interestingly, the benefit of spectral qubit ordering is most pronounced when combined with the TSN architecture, as we found that this particular combination produced the highest solution quality for both 3Reg and ER graphs. These findings indicate that improved performance in MPS-based simulations cannot be attributed solely to the logical qubit mapping, but rather emerges from the interplay between mapping strategy and logical qubit routing.

For the SK graphs we observed nearly identical performance between the RSN and TSN across all bond dimensions, except for $\chi = 128$ for which the TSN achieved a slightly lower error. A similar trend was also observed for ER graphs, for which the benefit of using spectral ordering and TSN only emerged at $\chi \geq 64$. These results suggest that the performance advantage associated with specific routing schemes may not persist for all bond dimensions. We hypothesize that the occasional benefit of the TSN arises because logical qubits experience smaller separations within this architecture, following more similar trajectories inside the TSN than in the RSN. Confirming this hypothesis would require a detailed analysis of the entanglement dynamics and correlations within the rectangular and triangular SWAP networks themselves, which we leave for future work.

Regarding the portfolio optimization problem, we found that the MPS-based solver produced solutions with a similar approximation ratio as on the MaxCut instances. These results are promising and suggest that the proposed optimization framework is well-suited not only for synthetic benchmark instances but also for realistic combinatorial optimization problems that exhibit complex structures and constraints.

In our analysis of the amount of entanglement exhibited by the MPS throughout portfolio optimization, we found that significantly larger amounts of entanglement were generated when using the hierarchical and spectral logical qubit orderings, as compared to the randomly shuffled ordering. Interestingly, these two orderings also resulted in the lowest error. While the analysis was performed on a single representative problem instance and therefore does not allow for strong general conclusions, it is nevertheless noteworthy that the instances exhibiting higher entanglement also achieved better approximation ratios. Rather than interpreting this as a causal relationship, it may be viewed as an indication that effectively utilizing the entanglement capacity of the MPS corresponds to a more expressive and better-optimized state representation. This interpretation aligns with recent findings in variational implementations of QITE ~\cite{morris_performant_2024}, where a similar analysis of the entanglement entropy suggested that the ability to harness and sustain nontrivial entanglement is an indicator of algorithmic performance. In this sense, the total entanglement generated during the optimization may serve as a useful metric for evaluating how effectively a given configuration of logical qubit mapping and routing explores highly correlated states within the Hilbert space.

\section{Conclusion}
In this work, we introduced an MPS-based imaginary-time evolution framework tailored for solving optimization problems with complex, non-local qubit connectivity. By combining spectral qubit mapping with rectangular and triangular SWAP networks built exclusively from local two-qubit gates, we demonstrated a scalable approach to simulating highly interconnected systems. The method was validated on both synthetic MaxCut instances of varying graph connectivity and on a dynamic portfolio optimization problem derived from real asset data.

Our results show that combining spectral qubit ordering with SWAP networks can substantially improve performance, in some cases reducing errors by more than an order of magnitude compared to shuffled qubit mappings. These improvements are consistently observed across both synthetic MaxCut benchmarks and a portfolio optimization problem based on real asset data. Notably, the specific combination of spectral qubit ordering and triangular SWAP networks proves particularly effective in finding low-energy solutions. Overall, the results emphasize that the benefits of optimized qubit mapping cannot be assessed in isolation but depend strongly on the structure of the routing network employed during time evolution. Furthermore, an analysis of entanglement entropy exhibited by the MPS during portfolio optimization revealed that spectral mapping produced higher and more evenly distributed entanglement across the bonds of the MPS. This suggests that better utilization of the MPS capacity of representing quantum correlations may correlate with improved optimization performance, though further study is required to confirm this link.

In future work, several directions appear promising for extending the proposed methods to a broader class of problems. One natural extension is to investigate real-time evolution in place of imaginary time evolution, which would enable the study of the dynamics generated by complex Hamiltonians with non-local structure. A natural problem to consider would be the simulation of an Ising model with long-range interactions, for which MPS-based methods have been proposed in the past~\cite{hauke_spread_2013, koffel_entanglement_2012, saadatmand_phase_2018}. Another possible extension is to increase the local physical dimension of the MPS sites beyond $d=2$, allowing integer-valued degrees of freedom to be represented directly and eliminating the need for binary encoding. Such an integer encoding would be relevant for applications such as portfolio optimization, but also for a broader class of graph-coloring style problems~\cite{jansen_qudit-inspired_2024, angkhanawin_graph_2025, deller_quantum_2023}. From a computational standpoint, recent advances have demonstrated that non-overlapping two-qubit gates can be applied to an MPS in parallel, achieving polynomial speedups and allowing for large-scale simulations exceeding 1000 qubits across hundreds of computational nodes~\cite{sun_improved_2024}. Incorporating similar parallelization strategies offers a promising route to bring MPS-based simulations, such as those developed in this work, into the high-performance computing (HPC) regime.

In summary, the results demonstrate that optimized logical qubit mapping and routing play an important role in extending the applicability of MPS-based methods for quantum simulation and optimization. By combining these algorithmic ideas with emerging parallelization strategies, MPS-based methods may evolve into a scalable framework for tackling increasingly complex quantum and classical problems.

\section{Code availability}
The code used to produce the numerical results presented in this work is openly available at \cite{swapnet_mps_ite}.

%\printbibliography
%\bibliography{references}

\bibliographystyle{quantum}
\bibliography{references_final}

\clearpage
\onecolumn
\appendix
\setcounter{equation}{0}
\renewcommand{\theequation}{A\arabic{equation}}

\section{Canonical forms of matrix product states}
\label{appendix: canonical form}
\subsection{Left- and right-canonical forms}
One of the main benefits of performing quantum simulations using MPS is the existence of so-called \textit{canonical forms}~\cite{vidal_efficient_2003}. Consider the quantum state
\begin{equation}
    \ket{\psi} = \sum_{i_1, i_2, ..., i_{N}} \psi_{i_1 i_2 \dots i_N}\ket{i_1 i_2 \dots i_{N}},
\end{equation}
with amplitudes encoded in the rank-$N$ tensor $\psi_{i_1 i_2 \dots i_N}$. To bring the state into canonical form, we successively reshape, decompose, and absorb singular values via singular value decompositions (SVDs).  

Bipartitioning between the first site and the remainder gives
\begin{equation}
     \psi_{i_1 i_2 \dots i_N} 
     \;\overset{\text{reshape}}{=}\; \psi_{i_1 (i_2\dots i_N)} 
     \;\overset{\text{SVD}}{=}\; \sum_{\alpha_1} U^{[1]}_{i_1 \alpha_1}\,\Lambda_{\alpha_1}\,V^\dagger_{\alpha_1 (i_2 \dots i_N)},
\end{equation}
where $U^{[1]}$ is left-orthogonal and $V^\dagger$ is right-orthogonal. Absorbing the singular values into the right tensor defines
\begin{equation}
    \psi_{i_1 (i_2\dots i_N)} = \sum_{\alpha_1} U^{[1]}_{i_1 \alpha_1}\, \psi'_{\alpha_1 i_2 \dots i_N}.
\end{equation}
Repeating the same procedure for the next bipartition,
\begin{equation}
    \psi'_{\alpha_1 i_2 \dots i_N} 
    \;\overset{\text{reshape}}{=}\; \psi'_{(\alpha_1 i_2)(i_3\dots i_N)}
    \;\overset{\text{SVD}}{=}\; \sum_{\alpha_2} U^{[2]}_{\alpha_1 i_2 \alpha_2}\,\Lambda_{\alpha_2}\,V^\dagger_{\alpha_2 (i_3\dots i_N)},
\end{equation}
and contracting the singular values again yields
\begin{equation}
    \psi'_{\alpha_1 i_2 \dots i_N} = \sum_{\alpha_2} U^{[2]}_{\alpha_1 i_2 \alpha_2}\, \psi''_{\alpha_2 i_3 \dots i_N}.
\end{equation}
Iterating this process site by site, the state factorizes as
\begin{equation}
    \psi_{i_1 i_2 \dots i_N} = 
    \sum_{\alpha_1, \alpha_2, \dots, \alpha_{N-1}}
    U^{[1]}_{i_1 \alpha_1}
    U^{[2]}_{\alpha_1 i_2 \alpha_2}
    \cdots
    U^{[N]}_{\alpha_{N-1} i_N},
    \label{eq: MPS left canonical form}
\end{equation}
which we call the \textit{left-canonical form} of the MPS, since each tensor is left-orthogonal.  

If the same decomposition is instead performed analogously from right to left, starting with site $i_N$, one obtains the \textit{right-canonical form}, consisting of right-orthogonal tensors:
\begin{equation}
     \psi_{i_1 i_2 \dots i_N} =
     \sum_{\alpha_1, \alpha_2, \dots, \alpha_{N-1}}
     V^{\dagger[1]}_{i_1 \alpha_1}
     V^{\dagger[2]}_{\alpha_1 i_2 \alpha_2}
     \cdots
     V^{\dagger[N]}_{\alpha_{N-1} i_N}.
     \label{eq: MPS right canonical form}
\end{equation}

\subsection{Mixed canonical form}
In addition to left- or right-canonical forms, it is often useful to bring an MPS into a \textit{mixed canonical form}. In this representation, all tensors to the left of some site $k$ are left-orthogonal, all tensors to the right of site $k$ are right-orthogonal, and the tensor at site $k$ carries the remaining normalization. This tensor is referred to as the \textit{orthogonality center}. Explicitly, the state can be written as
\begin{equation}
    \psi_{i_1 i_2 \dots i_N} = 
    \sum_{\alpha_1, \dots, \alpha_{N-1}}
    U^{[1]}_{i_1 \alpha_1}
    U^{[2]}_{\alpha_1 i_2 \alpha_2}
    \cdots
    U^{[k-1]}_{\alpha_{k-2} i_{k-1} \alpha_{k-1}}
    \, \Theta^{[k]}_{\alpha_{k-1} i_k \alpha_k} \,
    V^{\dagger[k+1]}_{\alpha_k i_{k+1} \alpha_{k+1}}
    \cdots
    V^{\dagger[N]}_{\alpha_{N-1} i_N}.
    \label{eq: MPS mixed canonical form}
\end{equation}
Here the tensors $U^{[j]}$ are left-orthogonal, the tensors $V^{\dagger[j]}$ are right-orthogonal. The central tensor at site $k$ can easily be related to the tensors at the same site in the left- and right-canonical forms as
\begin{equation}
    \Theta^{[k]}_{\alpha_{k-1} i_k \alpha_k} = U^{[k]}_{\alpha_{k-1} i_k \alpha_k} \Lambda_{\alpha_k} = \Lambda_{\alpha_{k-1}} V^{\dagger[k]}_{\alpha_{k-1} i_k \alpha_k}
    \label{eq: theta to U to V}
\end{equation}
We note that the right- and left-canonical forms are therefore just special cases of the mixed canonical form, where the orthogonality center has been positioned at the first and last site, respectively.

The mixed canonical form is particularly convenient because it makes expectation values and overlaps simple to compute, as tensor contractions to the left and right of the orthogonality center yield identities due to the orthogonality conditions, leaving only the local contraction at site $k$.

\subsection{Moving the orthogonality center}
The orthogonality center of the mixed canonical form~(\ref{eq: MPS mixed canonical form}) can be shifted one site to the right as follows. First, we reshape the current orthogonality center $\Theta^{[k]}_{\alpha_{k-1} i_k \alpha_{k}}$ on site $k$ by grouping the indices $\alpha_k$ and $i_k$. Next, we perform an SVD $\Theta^{[k]}_{(\alpha_{k-1} i_k) \alpha_{k}} = U^{[k]}_{(\alpha_{k-1} i_k)\beta_k}\Lambda_{\beta_k}W^\dagger_{\beta_k \alpha_k}$ followed by a contraction of $\Lambda_{\beta_k}W^\dagger_{\beta_k \alpha_k}$ onto the right-orthogonal tensor at site $k+1$ in order to shift the orthogonality center one site to the right. The whole procedure is as follows
\begin{equation}
\begin{aligned}
\sum_{\alpha_k}
\Theta^{[k]}_{\alpha_{k-1} i_k \alpha_{k}}\,
V^{\dagger[k+1]}_{\alpha_k i_{k+1} \alpha_{k+1}}
&\overset{\text{reshape}}{=}
\sum_{\alpha_k}
\Theta^{[k]}_{(\alpha_{k-1} i_k)\,\alpha_{k}}\,
V^{\dagger[k+1]}_{\alpha_k i_{k+1} \alpha_{k+1}}
\\
&\overset{\text{SVD}}{=}
\sum_{\beta_k,\alpha_k}
U^{[k]}_{(\alpha_{k-1} i_k)\,\beta_k}\,
\Lambda_{\beta_k}\,
W^\dagger_{\beta_k \alpha_k}\,
V^{\dagger[k+1]}_{\alpha_k i_{k+1} \alpha_{k+1}}
\\
&=
\sum_{\beta_k}
U^{[k]}_{\alpha_{k-1} i_k \beta_k}\,
\left(
\sum_{\alpha_k}
(\Lambda W^\dagger)_{\beta_k \alpha_k}\,
V^{\dagger[k+1]}_{\alpha_k i_{k+1} \alpha_{k+1}}
\right)
\\
&=
\sum_{\beta_k}
U^{[k]}_{\alpha_{k-1} i_k \beta_k}\,
\Theta^{[k+1]}_{\beta_k i_{k+1} \alpha_{k+1}} .
\end{aligned}
\end{equation}
The orthogonality center can analogously be moved one site to the left by instead contracting $U^{[k]}_{(\alpha_{k-1} i_k)\beta_k}\Lambda_{\beta_k}$ with the left-orthogonal tensor on site $k-1$. To move the orthogonality center multiple sites, the procedure is simply repeated recursively. In terms of time complexity, both the SVD and the following tensor contraction scale as $\mathcal{O}(d\chi^3)$.

\setcounter{equation}{0}
\renewcommand{\theequation}{B\arabic{equation}}
\section{Applying local two-qubit gates onto matrix product states}
\label{appendxi: 2-qubit gate}
Two-qubit gates acting on adjacent MPS sites can be efficiently applied by exploiting the canonical form of the MPS. Let $G_{i_{k}' i_{k+1}' i_k i_{k+1}}$ be a rank-4 tensor representing a two-qubit gate acting on the adjacent sites $(k,k+1)$. Before applying the gate, the orthogonality center of the MPS is shifted to one of these sites, as this will allow us to restore the canonical form using a single SVD after the gate has been applied. The gate is then contracted with the corresponding physical indices of the MPS, and an SVD is performed to restore the canonical form. Formally, this reads
\begin{equation}
\begin{split}
\sum_{i_k, \alpha_k, i_{k+1}} \Theta^{[k]}_{\alpha_{k-1} i_k \alpha_k} 
\, V^{\dagger[k+1]}_{\alpha_k i_{k+1} \alpha_{k+1}} 
\, G_{i_k' i_{k+1}' i_k i_{k+1}}
&\overset{\text{contract}}{=} T_{\alpha_{k-1}, i_k', i_{k+1}', \alpha_{k+1}} \\
&\overset{\text{SVD}}{=} \sum_{\alpha_k} 
U^{[k]}_{\alpha_{k-1} i_k' \alpha_k} \Lambda_{\alpha_k}' 
V^{\dagger[k+1]}_{\alpha_k i_{k+1}' \alpha_{k+1}} \\
&= \sum_{\alpha_k} \tilde{\Theta}^{[k]}_{\alpha_{k-1} i_k' \alpha_k} 
\tilde{V}^{\dagger[k+1]}_{\alpha_k i_{k+1}' \alpha_{k+1}},
\end{split}
\label{eq:apply-2qubit-gate}
\end{equation}
where $\tilde{\Theta}^{[k]}$ and $\tilde{V}^{\dagger[k+1]}$ denote the updated tensors after applying the gate. Since the canonical center was positioned at the sites on which the gate acts, the canonical form is automatically restored after the SVD. For an MPS with bond dimension $\chi$ and physical dimension $d$, the dominant cost comes from the SVD, giving a time complexity of $\mathcal{O}(d^3 \chi^3)$.

\end{document}